\documentstyle[12pt]{article}

\title{ Chiral  phase dependence of fermion partition function
in two dimension}

\author{
 Hisashi Kikuchi
\vspace*{0.6em}\\
{\normalsize \sl Department of Physics}\\
{\normalsize \sl University of California, Riverside}\\
{\normalsize \sl Riverside, CA 92521}
}

\date{June, 1993, UCRHEP-T110}

\begin{document}

\maketitle

\vfil

\begin{abstract}
The chiral phase dependence of fermion partition function in
spherically symmetric U(1) gauge field background
is analyzed in two dimensional space-time.
A well-defined method to calculated the path integral which apply to the
continuous fermion spectrum is described.
The one-to-one correspondence between the nonzero energy continuous spectra of
two pertinent hamiltonians, which are defined by the Dirac operator
to make the path integral well-defined,  is shown to be exact.
The asymptotic expansion for the chiral phase dependence in
\(1/|m|^2\) (\(m\) is mass of the fermion.) is proposed and the
coefficients in the expansion are evaluated up to the next-to-next
leading term.
Up to this order, the chiral phase dependence is given only by the winding
number of background field and the corrections vanish.
\end{abstract}

\newpage

\newcommand{\rpOf}{\mbox{Re}\,}
\newcommand{\ipOf}{\mbox{Im}\,}

\newcommand{\be}{\begin{equation}}
\newcommand{\ee}{\end{equation}}
\newcommand{\bea}{\begin{eqnarray}}
\newcommand{\eea}{\end{eqnarray}}

\newcommand{\del}[2]{\delta (  #1 -  #2 )}

\newcommand{\R}[3]{ R ( #1;  #2,  #3 ) }
\newcommand{\K}[3]{ K ( #1;  #2,  #3 ) }
\newcommand{\Ano}{{\cal A}}
\newcommand{\A}[1]{ {\cal A} ( #1)}
\newcommand{\SD}[2]{\rho ( #1;  #2 )}

\newcommand{\Rp}[3]{ R_+ ( #1;  #2,  #3 ) }
\newcommand{\Kp}[3]{ K_+ ( #1;  #2,  #3 ) }
\newcommand{\Rm}[3]{ R_- ( #1;  #2,  #3 ) }
\newcommand{\Km}[3]{ K_- ( #1;  #2,  #3 ) }
\newcommand{\Rpm}[3]{ R_\pm ( #1;  #2,  #3 ) }
\newcommand{\Kpm}[3]{ K_\pm ( #1;  #2,  #3 ) }
\newcommand{\SDp}[2]{\rho_+ ( #1;  #2 )}
\newcommand{\SDm}[2]{\rho_- ( #1;  #2 )}
\newcommand{\SDpm}[2]{\rho_\pm ( #1;  #2 )}
\newcommand{\SAS}[2]{\Delta \rho( #1, #2 )}

\newcommand{\Ddagger}{D^\dagger}
\newcommand{\Hp}{H_+}
\newcommand{\Hm}{H_-}
\newcommand{\Hpm}{H_\pm}
\newcommand{\phip}{\phi_+^{(n)}}
\newcommand{\phim}{\phi_-^{(n)}}
\newcommand{\psip}{\psi_+{}}
\newcommand{\psim}{\psi_-{}}

\newcommand{\rsol}[3]{\psi^{#1}(#2, #3)}
\newcommand{\freeu}[3]{u^{#1}(#2,#3)}
\newcommand{\freew}[3]{w^{#1}(#2,#3)}
\newcommand{\Reg}[3]{ \varphi^{#1}( #2, #3) }
\newcommand{\Jost}[3]{ f^{#1}(#2,#3) }

\newcommand{\absn}{{|n|}}
\newcommand{\Jostf}[2]{ F^{#1}(#2) }
\newcommand{\WronskianOf}[2]{ W\left\{ {#1}, {#2} \right\} }
\newcommand{\phshift}[2]{ \delta^{#1}(#2) }
\newcommand{\radialR}[4]{ R^{#1}( #2; #3, #4) }
\newcommand{\rx}{r_x}
\newcommand{\ry}{r_y}

\setlength{\baselineskip}{18pt}

\section{Introduction}

The quark partition function \(Z\) in given gluon background has
nontrivial dependence on chiral phase \(\alpha\),
\be \alpha = \mbox{arg }\det m, \ee
of the mass matrix \(m\)
\cite{Jackiw.et.al+Callan.et.al:1976,Peccei.et.al:1977,%
Weinberg+Wilczek:1978}.
The \(\theta\)-term in QCD lagrangian comes from this  \(\alpha\)-dependence
and leads us to strong CP problem
\cite{Jackiw.et.al+Callan.et.al:1976,Peccei.et.al:1977,%
Weinberg+Wilczek:1978,Baluni+:1979}.

In the previous work \cite{Kikuchi:1992}, we have evaluated the
\(\alpha\)-dependence, \(d\ln Z/d\alpha\), by
calculating functional Jacobian in fermionic path integral
\cite{Fujikawa:1979,Fujikawa:1984}.
The path integral is calculated by solving eigenstates and eigenvalues
of two pertinent hamiltonians, \(\Hp\) and \(\Hm\),
which is constructed from the Dirac operator
of the model.
Applying a box quantization to each eigenstates,
we discretised the spectrum and have evaluated the Jacobian in a manifest
fashion.
An important thing which was took for granted in the previous calculation
 was  the existence of the one-to-one correspondence
between nonzero eigenstates for \(\Hpm\):
the spectrum of \(\Hp\) is the same as that of \(\Hm\) except for
the zero modes.
Under the existence of the correspondence,
the contribution to \(d\ln Z/d\alpha\) from the nonzero modes
cancels and only zero mode contribution remains.
As a result we get only the \(\theta\)-term  and no other correction
terms for \(d\ln Z/d\alpha \).

In a certain background configuration, the box quantization is not
necessarily possible.
In this case, we need to start with the infinite size system and deal
with spectrum with  continuous  portion from the beginning.
The one-to-one correspondence  then  becomes vague.
Indeed, there have been various intensive studies on this point, especially
related to calculation of the index of  Dirac operators
\cite{Kiskis+Jackiw.et.al+Ansourian,Callias:1978+Bott.et.al:1978%
,openA,openB,Niemi.et.al:1984,Akhoury.et.al:1986,Niemi.et.al}.
It has been pointed out that in some models the one-to-one correspondence
is no longer hold \cite{openA,openB}:
the spectral densities of \(\Hpm\) for the continuous portion are different.
If the one-to-one correspondence is violated, \(d\ln Z/d\alpha\)
has extra contribution from the states which belongs to the continuum.

The purpose of this paper is  to develop the calculation of the
\(\alpha\)-dependence in a well-defined fashion which apply for the case that
the spectrum contains  continuous portion
and to see if there is any correction in
the \(\alpha\)-dependence of the partition function.
We will work in a model of a fermion coupled to
U(1) gauge field \(A_\mu\) in two dimension.
As for chiral property, the fermion  shares
similarity with quarks in QCD: the axial U(1) anomaly is given by the
winding number density of the background field.
Thus we hope the technique developed in this paper also apply for QCD.

We specifically work for spherically-symmetric  background field.
This restriction enables us to analyze the problem in terms of the
scattering theory for a particle in a centrifugal potential in two dimension.
The spectrum is continuous and extends to zero energy.
We consider the background as having any real winding number
\be \nu \equiv {e\over 2\pi} \int F_{12} d^2\!x,
\quad F_{12} = \partial_1 A_2 - \partial_2 A_1, \label{eq:wind}\ee
which prevents us from compactifying  the space-time and
from discretising the spectrum.
Depending on whether the  winding number is zero or not,
the  potential possesses \( 1/r^2 \)-tail at spatial infinity.
This fact causes  non-analyticity of the scattering waves at zero energy
and will turn out essential for nontrivial \(d\ln Z/d\alpha\).

This paper is organized as follows. In section 2, we review the calculation
for \(d\ln Z/d\alpha\) for  discrete spectrum
and rewrite it in terms of spectral
density.  This form of \(d\ln Z/\alpha \) enables us to calculate it
for the continuous spectrum.
In section 3, we describe the definitions of  spectral densities and
write \(d\ln Z/d\alpha\) in two terms,
a contribution from nonzero energy continuous spectrum and
those from zero energy states.
We use contour integral to  regularize the zero energy
contribution.
In section 4, we define Jost functions in two dimensional scattering problem
and write the spectral density in terms of them.
In section 5, we show  the one-to-one correspondence for the nonzero energy
states  is exact.
In section 6, we will evaluate  the zero energy contributions.
We adopt the asymptotic expansion of \(d\ln Z/d\alpha \) in the inverse
of squared mass parameter \(1/|m|^2\) and calculate the coefficients
up to the second order term.
Section 7 is devoted for summary and discussion.

\section{The \(\alpha\)-dependence of Z}
We take a fermion \(\psi\) in two dimension with a complex mass \(m\) and
coupling to external U(1) gauge field \( A_\mu\).
The lagrangian in Euclidean metric is of the form
\be  L = (-i)  \bar\psi \left[ (\partial_\mu -ie A_\mu )
\sigma_\mu - m \left( {1+\sigma_c \over 2}\right)
- m^* \left( {1 - \sigma_c \over 2 }\right)  \right] \psi, \label{eq:L}
\ee
where \(e \) is the coupling constant, \(\sigma_\mu\) (\(\mu = 1,2\))
 are the first and second
components of Pauli matrices,
\be \sigma_1 = \left( \begin{array}{cc} 0 & 1 \\ 1 & 0 \end{array}
\right), \quad
\sigma_2 = \left( \begin{array}{cc}  0 & -i  \\ i  & 0 \end{array}
\right); \ee
the third component is used to define chirality,
\be \sigma_c \equiv \sigma_3 = \left( \begin{array}{cc} 1 & 0 \\ 0 & -1
\end{array} \right). \ee
The chiral phase \(\alpha\) is simply related to \(m\) as
\be m = |m| e^{i\alpha}. \ee

We intend to examine the \(\alpha\)-dependence of the partition function
\be Z \equiv \int [d\psi][d\bar\psi] e^{ - \int d^2\!x L }. \label{eq:defZ}\ee
Specifically we evaluate the derivative \(d\ln Z/d\alpha\).
By definition, it is given by
\bea {d \ln Z \over d\alpha} & = & {1\over Z} \int [d\psi][d\bar\psi]
\int d^2\!x \left(-{dL\over d\alpha}\right) e^{- \int d^2\!x L }
\nonumber \\
& = &
\int d^2\!x \left\langle \bar\psi \left[ m  \left({1+ \sigma_c \over 2}\right)
 - m^*  \left({1-  \sigma_c \over 2}\right) \right] \psi \right\rangle,
\label{eq:CD1}\eea
where \(\langle O\rangle\) denotes the expectation value of operator \(O\)
in  a given gauge field background.

The evaluation of \(Z\) and \(d\ln Z/d\alpha\) in terms of fermionic
eigenstates in gauge field background has been given by
discretising the spectrum with box quantization \cite{Kikuchi:1992}.
Let us briefly review the results.
The Dirac operator \(\cal L\) has an explicit form
\be {\cal L } = \left( \begin{array}{cc} -im & D \\ \Ddagger & -im^*
\end{array}
\right) \ee
with
\bea D & \equiv& i (\partial_1 - ieA_1) + (\partial_2 - ie A_2 ),
\label{eq:2.8}\\
\Ddagger &\equiv&  i (\partial_1 - ieA_1) - (\partial_2 - ie A_2 ),
\label{eq:2.9}\eea
for lagrangian (\ref{eq:L}).
It is not hermitian for complex \(m\) and thus
we cannot use eigen modes of \(\cal L\) to define the functional space.
Instead, we use eigenstates  \(\phip \) and \(\phim \)
of ``hamiltonians''
\bea \Hp & \equiv & D \Ddagger
= - \left(\partial_\mu - ie A_\mu\right)^2 - e F_{12},
\\
\Hm & \equiv & \Ddagger D =
- \left(\partial_\mu - ie A_\mu\right)^2 + e F_{12},
\label{eq:Hpm} \eea
where \(n\) is integer and used to label different eigenstates.
The eigenvalues \(E_\pm^{(n)}\) are non-negative.
The spectrum of \(\Hpm\) coincide for nonzero eigenvalues.
This can be seen  by noticing the one-to-one correspondence between
\(\phip\) and \(\phim\),
\bea
\phim  & = & {1\over \sqrt {E^{(n)}}} \Ddagger \phip , \label{eq:susy1}\\
\phip  & = & {1\over \sqrt {E^{(n)}}} D \phim , \label{eq:susy2}\eea
where \(E^{(n)} \equiv  E_+^{(n)} =  E_-^{(n)}\).

Expanding the field \(\psi\) and \(\bar\psi\) in terms of these eigenstates,
diagonalising the action by an variable transformation for Grassmann
path-integral variable, and taking into account the corresponding Jacobian,
we get \cite{Kikuchi:1992}
\be Z = {\cal J} |m|^{ n_+ + n_- } \prod{}' \left( E^{(n)} + |m|^2 \right),
\label{eq:Z1} \ee
where \(n_\pm\) are the number of zero modes for \(\Hpm\) and
the prime on the product means the product must be taken over each pair of
nonzero eigenvalue.
\(\cal J\) denotes the Jacobian in the path-integral measure
and it carries the \(\alpha\)-dependence
as \(e^{i\alpha( n_+ - n_- ) } \).
Further the expectation value in (\ref{eq:CD1}) is given by
\be {d \ln Z  \over d\alpha}  =
 i \sum_n \left[ {|m|^2 \over E_+^{(n)} + |m| ^2 }
 - {|m|^2 \over E_-^{(n)} + |m| ^2 } \right]. \label{eq:CDt}\ee
Because of the one-to-one correspondence (\ref{eq:susy1})--(\ref{eq:susy2}),
the nonzero eigenvalues  cancel out and (\ref{eq:CDt}) is consistent
with the \(\alpha\)-dependence of \(\cal J\).

We extend these results to the case of continuous spectrum by generalizing
them in terms of the spectral density, the number density of states per
energy.
For the discretised spectrum, it is obviously given by
\be \rho_\pm (E)  = \sum_n \delta( E - E_\pm^{(n)} ). \label{eq:primary}\ee
We rewrite (\ref{eq:Z1}) and (\ref{eq:CDt}) as
\be \ln Z = \ln {\cal J} +
\int_0^\infty dE\, {1\over 2} \left[\rho_+ (E) + \rho_-(E) \right]
 \ln (E + |m|^2 )  \label{eq:Z} \ee
and
\be {d \ln Z \over d\alpha} = i\int_0^\infty dE
{|m|^2 \over E + |m| ^2 }
\left[ \rho_+( E ) - \rho_-( E ) \right]. \label{eq:CDZ}\ee
We take (\ref{eq:Z}) and (\ref{eq:CDZ}) as definitions for \(Z\) and its
\(\alpha\)-dependence in terms of spectral density \(\rho_\pm\);
we develop the way to calculate \(\rho_\pm\) that is applicable to  the
continuous spectrum.

We need to mention two problems about the definitions
(\ref{eq:Z})--(\ref{eq:CDZ}).
First,  we have to be careful in the \(E\)-integrals at \( E = 0\) to get
well-defined results
when the continuous portion in the spectrum extends to zero energy.
A more careful definition  which prevent the problem will be given
in the next section.

Secondly, the spectral density behaves as
\(\rho_\pm(E) \sim V/4\pi \) as \( E \rightarrow \infty\)
with the volume \(V\) of quantization box in two dimension.
Thus the \(E\)-integral in \(\ln Z\) has UV divergence.
We can define once-subtracted spectral densities  \(\delta\rho_\pm(E)\)
which is defined as the deviation of \(\rho(E)\) from the spectral density
 \(V/4\pi\) of the free hamiltonian.
In \(\ln Z\) we use \(\delta\rho_\pm(E)\) instead of \(\rho_\pm(E)\).
The perturbative calculation for \(\ln Z\), which measures its  deviation
from the case of free fermion, reveals that the correction to \(\ln Z\) is
finite \cite{Schwinger-model}.
Thus the one subtraction suffices to make \(\rho_\pm(E) \)
and \(\ln Z \) finite.
As for \(d\ln Z/d\alpha\), the divergence cancels between
\(\rho_+\) and \(\rho_-\) and no ill-definedness takes place.

\section{Spectral density}

The problem is now to develop a decent method for evaluation of
the spectral densities \(\rho_\pm(E) \).
The various methods have been intensively studied in connection
with fermion number fractionization and with Witten index in supersymmetric
quantum mechanics \cite{openB,Niemi.et.al:1984,Akhoury.et.al:1986,Niemi.et.al}.
A subtlety of the calculations related to the order of various
integrations are emphasized for the Dirac operators whose continuous portion
extend to zero energy \cite{openB}.
Here, we  adopt a different way of defining spectral densities.
The advantage of our description  is that it naturally leads us how
to regularize the contribution from the zero energy states.

We start with  heat kernel \( \K \tau x y \) for \(H\).
(We suppress the subscripts \(_\pm\) until we
need to distinguish \(\Hpm\).)
It is defined by  the  equation
\be \left( {\partial \over \partial \tau } + H \right) \K \tau xy = 0
\quad\mbox{for  } \tau \ge 0 \label{eq:K1}\ee
and the initial condition
\be \K 0 xy = \del x y, \label{eq:K2} \ee
where \(\del xy \) denotes the two dimensional Dirac \(\delta\)-function.
A set of  (\ref{eq:K1}) and (\ref{eq:K2}) fixes \(\K \tau xy \) completely.

The resolvent \( \R E xy \) is then  defined as the Laplace transform of
the heat kernel \( \K \tau xy \) by
\be \R E xy \equiv \int_0^\infty d\tau e^{E\tau} \K \tau xy.
\label{eq:R1} \ee
Using the definitions (\ref{eq:K1})--(\ref{eq:R1}),
we see \(\R E xy\) satisfies
\be \left( H - E \right) \R Exy = \del xy  \label{eq:R2}\ee
as long as the integral in (\ref{eq:R1}) converges.
According to general properties of Laplace transformation,
\(\R Exy \) for fixed \( x\) and \( y\) is a holomorphic function
of complex \( E \) whose real part is less than some critical value
\(E_c\), (\( \rpOf E < E_c \)).
In our case, the spectrum of \(H\) is nonnegative thus \(E_c = 0\).
We analytically continue \(\R Exy\) in a complex \(E\)-plane with a cut
along the real positive \(E\)-axis.

Obviously, \( \K \tau xy \) is restored by the inverse Laplace transformation
\be \K \tau xy = {1\over 2\pi i} \int_C dE e^{ - E\tau } \R Exy,
\label{eq:KinLa}\ee
where the integral contour C runs from \(c  - i\infty\) to \(c + i\infty\),
(\(c<0\)).

We define a state density  \( \SD Ex \) of \( H \),
the number of states per unit energy and unit space volume,
by the  discontinuity
of the  resolvent \( \R E xy \) at the cut \cite{Chadan.et.al:1989},
\be \SD Ex \equiv \lim_{y \rightarrow x} \lim_{ \epsilon \rightarrow +0}
{1\over 2\pi i} \left[ \R {E+i\epsilon}xy - \R {E -i\epsilon } xy \right].
\label{eq:dens}\ee
The order of two limitations in (\ref{eq:dens}),
(\(\epsilon \rightarrow +0 \)) before (\(y \rightarrow x \)),
cannot be altered.
This  circumvents   ill-definedness which happens in the space with
dimensionality larger than one.
In appendix \ref{app:consistency}, we give simple examples for
free hamiltonians in various dimensions which show the above definition
is consistent with our understandings to \(\SD Ex\).

The spectral density \(\rho(E)\)
is then defined formally by
\be \rho(E) \equiv \int d^2\!x \SD Ex, \ee
and the once-subtracted spectral density by
\be \delta\rho(E) \equiv \int d^2\!x \left[ \SD Ex - {1\over 4\pi}\right],
\label{eq:OS}\ee
where \(1/4\pi \) is the state density for the free hamiltonian.
This subtraction make the \(E\)-integral in (\ref{eq:Z}) finite.

Now let us turn into a regularization problem at \(E = 0\).
In the background we consider, the resolvents are singular there.
There are two reasons for the singularity:
the existence of normalizable zero modes \(\phi^{(0)}\) adds pole-type
singularity
\be \R Exy \sim {1\over E }\phi^{(0)}(x) \phi^{(0) \dagger} (y) \ee
to the resolvents;
since the continuous portion in the spectrum extends to zero energy,
the resolvents may have other type of  singularities as well.
The definition (\ref{eq:dens}) is not sufficient to take into
account these singularities.

To answer this problem,  let us recall the calculation of the anomaly,
the anomalous term in Ward-Takahashi identity.
According to Fujikawa, it is given by taking a functional trace of chirality
\(\sigma_c\) using the heat kernel regularization \cite{%
Fujikawa:1979,Fujikawa:1984}.
It corresponds to \(\tau\rightarrow 0\)
limit of
\be \Ano (\tau, x) \equiv \lim_{ y \rightarrow x  }
\left[ \Kp \tau xy - \Km \tau xy \right]. \label{eq:ano}\ee
[The limit (\( y \rightarrow  x \)) should be taken before (\(\tau
\rightarrow +0 \)),
otherwise we get trivially zero as the result because of
the initial condition (\ref{eq:K2}).]
This quantity \(\lim_{\tau\rightarrow 0} \Ano(\tau, x) \) must be
the difference between the state densities per unit spatial volume
for \(\Hp\) and \(\Hm\):
we use eigenstates of \(\Hpm\) for the basis in the functional space
and they have definite chirality \(\pm\); thus the difference
between the state densities generates nonzero Jacobian in the path integral
measure under a chiral transformation \(e^{i\theta(x) \sigma_c }\)
\cite{Fujikawa:1979,Fujikawa:1984}.
We use the Laplace transform (\ref{eq:KinLa}) for the heat kernels
in (\ref{eq:ano}) and
bend the integral contour C so that it pinches
real positive \(E\) axis leaving a small circle C\('\) around the origin,
to get
\bea \Ano(\tau, x) & = & \lim_{  y \rightarrow  x}
{1\over 2\pi i} \oint_{C'} dE e^{-E\tau}
\left[ \Rp Exy - \Rm Exy \right] \nonumber \\
& & + \int_\epsilon^\infty dE e^{-E\tau} [ \SDp Ex- \SDm Ex ],
\label{eq:AinR}\eea
where C\('\) is a clockwise oriented with radius \(\epsilon\).
(See Fig.~1.)
The second term represents what it must represent:
it is the difference between the state densities contributed from
the nonzero energy states.
The first term tells us how to regularize the
contributions from zero energy states.
If the singularity comes from  normalizable zero modes,
the contour integral over C\('\) properly gives the density by
zero modes. The expression in (\ref{eq:AinR}), however,
applies to wider class of singularities.

Incidentally, note that the calculation of the anomaly in term of
 \(\Ano(\tau,x) \) is gauge invariant procedure:
two different heat kernels for \(A_\mu\) and its gauge transform \(A'_\mu
= A_\mu - \partial_\mu \Lambda / e \) are related by
\be \K \tau xy = e^{i (\Lambda(x) - \Lambda(y) ) } K'(\tau;x,y); \ee
the difference disappears after we take the limit (\(y\rightarrow x\)).

Applying the same reason to regularize the contributions from
zero energy states,  we finally adopt the definition\footnote{%
A similarly regulated expression for \(\ln Z\) in terms of
\(\delta\rho_\pm(E)\) is easily given.}
\bea
{d \ln Z \over d\alpha} & = & i
\int d^2\!x \lim_{  y \rightarrow  x}
{1\over 2\pi i} \oint_{C'} dE {|m|^2 \over E + |m| ^2}
\left[ \Rp Exy - \Rm Exy \right].
\nonumber\\
&&+ i\int_\epsilon^\infty dE
{|m|^2 \over E + |m| ^2 }
\left[ \rho_+( E ) - \rho_-( E ) \right].
\label{eq:ad}
\eea

\section{Jost function in two dimension}

We now apply the method described for spectral density in the previous
section to the case of spherically symmetric  background field.
The problem reduces to solving  Shr\"odinger equations of
two dimensional scattering.
According to  a well-established technique in three dimensional scattering
\cite{Chadan.et.al:1989},
we define two types of solutions for the Schr\"odinger equations,
regular solutions and Jost solutions, and define Jost function in terms of
them.
We can derive resolvents and  spectral densities in terms of them.
Especially we will see that  spectral density is related to the Jost
function.

Let us define  the polar coordinates \(r\) and \(\theta\)  by
\be x_1 = r \cos \theta, \quad x_2 = r \sin\theta. \ee
We take  radial gauge \(A_r = 0 \) and choose \(A_\theta\) which depends
only on \(r\), where
\bea
A_r & = &  A_1  \cos\theta + A_2  \sin \theta, \\
A_\theta & = & - A_1 r \sin\theta + A_2 r \cos \theta.
\eea
The resulting field strength \(F_{12} = d A_\theta /rdr \) is then
spherically symmetric.

The \(A_\theta\) must vanish at the origin  so that \(A_\mu\) is continuous.
To avoid any subtle singularity which may appear when we calculate derivatives
of \(A_\mu\) with respect to coordinates at \(r = 0 \), we further restrict
\(A_\theta\) to such a configuration that its derivatives of any order
with respect to \(r\)
vanish at \(r=0\). An simple example of such a configuration is
\(A_\theta \sim e^{- 1/r } \).

For the background to have nonzero winding number, \(A_\theta\) approaches
to a constant,
\be e A_\theta \sim \nu,
\label{eq:Atinfty} \ee
as \(r\) goes to infinity.
Note  the constant \(\nu \) is the winding number in (\ref{eq:wind}).

In this  background, the derivative operators (\ref{eq:2.8})--(\ref{eq:2.9})
are
\bea
D = ie^{-i\theta} \left( {\partial \over \partial r} - {i\over r}
 {\partial \over \partial \theta} - {e A_\theta \over r} \right),
\label{eq:4.5}\\
\Ddagger = ie^{i\theta} \left( {\partial \over \partial r} + {i\over r}
 {\partial \over \partial \theta} + {e A_\theta \over r} \right).
\label{eq:4.6}
\eea
The Schr\"odinger equations \(\Hpm \phi_\pm = E \phi_\pm\)
with \( E = k^2 \) read
\be
\left[ - { d^2 \over dr^2 } + {1\over r^2}\left( n^2 - {1\over 4} \right) +
V_\pm (r) \right] \psi_\pm^n(r) = k^2 \psi_\pm^n (r), \label{eq:scte}\ee
where \(\psi_+^n(r)\) and \(\psi_-^n(r)\) denote partial waves
with definite integer angular momentum \(n\)
\be \phi_\pm (x) = e^{in\theta} {1\over \sqrt r} \psi_\pm^n (r), \ee
and the potentials are
\be
V_\pm(r) = {1\over r^2 }\left( e^2 A_\theta^2 - 2 n e A_\theta \right) -(\pm)
{1\over r} {d ( e A_\theta ) \over dr }. \label{eq:potential} \ee
Since the procedure hereafter applies equally to both \(\psi_+^n\)
and \(\psi_-^n\), we suppress the subscript \(\pm\).

In background free case, \(V(r) = 0\), the typical two solutions of
(\ref{eq:scte}) are
\bea
\freeu nkr & = & \left\{ \begin{array}{l}
  {\displaystyle\sqrt{ \pi r \over 2 }} J_{n}(kr) \quad\mbox{for  } n\ge 0 \\
  (-)^{\absn}  {\displaystyle\sqrt{ \pi r \over 2 }} J_{\absn}(kr)
\quad\mbox{for } n< 0
\end{array} \right.\label{eq:u} \eea
and
\bea \freew nkr & = & \left\{ \begin{array}{l}
i {\displaystyle\sqrt{ \pi r \over 2} } H_{n}^{(1)}(kr)
\quad\mbox{for } n\ge 0 \\
 (-)^{\absn} i {\displaystyle\sqrt{ \pi r \over 2} }H_{\absn}^{(1)}(kr)
\quad\mbox{for  } n < 0  \end{array}\right., \label{eq:w}\eea
where \( J_n \)   and
\(H_n^{(1)}\) denote the Bessel function and the first kind Hankel function
of order \( n\), respectively.
The phase convention for \( n < 0\) in (\ref{eq:u})--(\ref{eq:w})
helps expressions below to be simpler.
(See, e.g.,  (\ref{eq:susy3})--(\ref{eq:susy4}).)
The free solutions  asymptotically approximate to
\bea
\freeu nkr & \sim  & {1\over \sqrt k}\cos\left( kr - { ( n + 1/2 ) \pi
\over 2 } \right), \label{eq:uati}\\
\freew nkr &\sim& {i\over\sqrt k} \exp\left[i\left( kr - { ( n + 1/2 ) \pi
\over 2 } \right) \right], \label{eq:wati}
\eea
and their Wronskian is
\be \WronskianOf{\freew nkr}{\freeu nkr} \equiv
\freew nkr {d\freeu nkr\over dr} - {d\freew nkr \over dr }
\freeu nkr = 1. \label{eq:Wron}\ee
Based on these free solutions, we define regular solution \(\Reg nkr\)
and Jost solution \(\Jost nkr\) by integral equations
\bea
\Reg nkr & = & \freeu nkr
- \int_0^r dr' \left[ \freew nkr \freeu nk{r'}\right.\nonumber\\
&& - \left.\freew nk{r'}\freeu nkr \right] V(r')\Reg nk{r'}
\label{eq:reg1}\eea
and
\bea
\Jost nkr & = & \freew nkr
+ \int_r^\infty dr' \left[ \freew nkr \freeu nk{r'}\right. \nonumber\\
&&- \left.\freew nk{r'}\freeu nkr \right] V(r')\Jost nk{r'}.
\label{eq:jost1}\eea
These equations are solved by iteration and defines a series for both
\(\varphi^n\) and \(f^n\).

The advantage of defining the solutions by the integral equations, or
the series expansions  generated by the equations,  is that
we can regard the solutions as analytic function
of complex values of \( k\) for fixed real positive \( r\).
By the same  estimation about convergence of the series as
in Ref.~\cite{Chadan.et.al:1989}, we can see the following analytic nature
of the solutions.
For \( \Reg nkr\), the series  absolutely converges if the potential
meets  a condition
\be \int_0^r dr' { r' |V(r')| \over 1 + |k| r' } < \infty. \label{eq:reg2}\ee
Then, since each term in the series is an entire function of \(k\),
\(\Reg nkr \) is also an entire function of \(k\).
[Strictly speaking the condition (\ref{eq:reg2}) is for \( n\neq 0\).
Note that \(H_n^{(1)}(kr)\) includes a term proportional to \( J_n(kr) \ln kr
\). For \( n = 0\),  this term requires that the potential
obeys a slightly stronger condition
\(\int_0^r  dr' r'\, |V(r')| \ln r'  < \infty \) at \(r \sim 0\).
Note also that the part proportional to \(\ln k\) cancel in (\ref{eq:reg1}) and
does not affect the entireness of the series with respect to \(k\).]
Similarly, the series for \(\Jost nkr\) absolutely converges if
\(\ipOf k \ge 0 \) and
\be \int_r^\infty dr' { r' |V(r')| \over 1 + |k| r' } < \infty.
\label{eq:jost2}\ee
\(\Jost nkr \) is, then,  a holomorphic function in upper half \(k\)-plane.

Note the potential (\ref{eq:potential}) with (\ref{eq:Atinfty})
has \(1/r^2 \)-tail  at infinity  for nonzero winding number \(\nu\).
Thus the point \( k = 0 \) is the singular point for the Jost solutions.

The regular solution and Jost solution satisfy some symmetry properties.
They are derived by noticing that
\( [\Reg nkr]^* \) and \( \Reg n{-k^*}r \) satisfy the same wave equation
and are both regular at \( r \sim 0\).
Thus  they should be proportional to each other.
The same reason applies to \(\Reg nkr\) and \(\Reg n{-k}r\).
By using the behavior at \(r \sim 0 \),
\be \Reg nkr \sim \freeu nkr  \sim {\sqrt \pi \over |n| !} k^\absn
\left({r\over 2}\right)^{(\absn + 1/2)},
\label{eq:Reg23}
\ee
we obtain
\be \left[ \Reg n k r\right]^* = (-)^\absn  \Reg n {-k^*}r \quad\mbox{and}\quad
\Reg nkr = (-)^\absn \Reg n{-k}r.
\label{eq:reg3}\ee
Similarly, noticing \(\Jost nkr \sim \freew nkr\) at  \( r \sim \infty \)
and (\ref{eq:wati}), we get
\be \left[\Jost nkr \right]^* = (-)^\absn \Jost n{-k^*}r .\label{eq:jost3}\ee

We define the Jost function in terms of Wronskian by
\be  \Jostf nk = \WronskianOf { \Jost nkr }{\Reg nkr }. \label{eq:jostf}\ee
\(\Jostf nk\) is a holomorphic function in upper
\(k\)-plane and obeys a condition
\be \Jostf n k \rightarrow 1 \quad \mbox{as  } |k| \rightarrow \infty \ee
for the potentials that meets condition (\ref{eq:reg2}) with \(r = \infty\)
\cite{Chadan.et.al:1989}.
Further by (\ref{eq:reg3}) and (\ref{eq:jost3}), we see
\be \left[ \Jostf n k \right]^* = \Jostf n { -k^*} .  \ee
\(\Reg nkr \) can be expressed in a linear combination of \(\Jost nkr\) and
a solution that is linearly independent from \(\Jost nkr\), say \(g^n(k,r)\).
The Jost function represents an amplitude of \(\Reg nkr\) with respect to
\(g^n(k, r)\) in the linear combination.
Especially,  for real positive \(k\),
\( \Jost n kr \) and \( \Jost n {-k} r\)
are the linearly independent two solutions.
The linear combination for \( \Reg nkr\) in this case is given by
\be \Reg n kr = {i\over 2} \left[ (-)^\absn \Jostf n k \Jost n {-k} r
- \Jostf n {-k} \Jost nkr \right]\quad \mbox{(real positive \(k\))}.
\label{eq:RbyJ} \ee

Now we can write the resolvent \(\R Exy \) using
\(\Reg nkr\), \(\Jost nkr\), and  \(\Jostf nk \)
in the following way.
First define a function \(\radialR n k r {r'}\) which is holomorphic in the
upper half \(k\)-plane by
\bea \radialR n k r {r'} &\equiv&
{ 1\over \Jostf n k } \left[ \theta( r - r') \Jost n kr \Reg n k {r'} \right.
\nonumber\\
&& \left.+ \theta ( r' - r) \Reg n kr \Jost n k {r'} \right]. \eea
\(\radialR nkr{r'}\) satisfies
\be
\left[ - { d^2 \over d r^2 } + { 1\over r^2 } \left ( n^2 - {1\over 4} \right)
+ V(r) - k^2 \right] \radialR n k r {r'}  = \delta( r -r' ). \label{eq:rR}\ee
Then \( \R E xy \) for \(E = k^2 \) is given by
\be \R Exy = {1\over 2\pi \sqrt{ \rx\ry } }
\sum_{n = -\infty}^\infty e^{i n (\theta_x - \theta_y )}
 \radialR n k {\rx}{\ry},  \ee
where \(\rx\) (\(\ry\)) is the polar coordinates for \( x\) (\(y\)).
The resolvent defined this way obeys (\ref{eq:R1}) and is holomorphic
in complex \(E\)-plane with a cut along the positive real axis.

One may wonder  an  ambiguity to add an linear combination of
terms made of the product of the solutions, like \( \Reg nkr \Jost nk{r'} \),
to \(\radialR nkr{r'}\). It  does not change
the equation (\ref{eq:rR}). Two requirements fix this ambiguity:
\(\R Exy \)  must be regular at both \(\rx \sim 0 \) and \(\ry \sim 0\)
since the background \(A_\mu\) is regular there; \(\R Exy \)  must be a
decreasing  function of \(|x - y |\) as \( | x - y | \rightarrow \infty \)
so is \(\K \tau xy \).

Remembering the definition for the state density (\ref{eq:dens}) and
using the relation (\ref{eq:RbyJ}) for real positive \(k\),
we get
\be \SD Ex  =  {1\over  2\pi^2 r } \sum_{n = -\infty}^{\infty}
{ \left[ \Reg n {\sqrt E} r \right]^2
\over \Jostf n{\sqrt E} \Jostf n{-\sqrt E} }.\label{eq:density}\ee
We further calculate the once-subtracted spectral density \(\delta\rho(E)\).
In appendix~\ref{app:density}, we show \( \delta\rho (E ) \)
is related to the Jost function by\footnote{%
The relation can be given  by using the equivalence
of Jost function and Fredholm determinant for the case of \(\Hpm\) being
Fredholm.
In the problem we are considering, \(\Hpm\) are {\em not}
Fredholm operators.}
\be \delta\rho (E) = { -1\over 2\pi i} \sum_{n=-\infty}^{\infty}
{1\over 2k}{d\over dk} \left[ \ln\Jostf nk - \ln\Jostf n{ - k } \right].
\label{eq:SDinJ}\ee

\newcommand{\Regp}[2]{ \varphi_+^{n}( #1, #2) }
\newcommand{\Regm}[2]{ \varphi_-^{n+1}( #1, #2) }
\newcommand{\Jostp}[2]{ f_+^{n}(#1,#2) }
\newcommand{\Jostm}[2]{ f_-^{n+1}(#1,#2) }
\newcommand{\Jostfp}[1]{ F_+^{n}(#1) }
\newcommand{\Jostfm}[1]{ F_-^{n+1}(#1) }
\newcommand{\radialRp}[3]{ R_+^{n}( #1; #2, #3) }
\newcommand{\radialRm}[4]{ R_-^{n+1}( #1; #2, #3) }

\newcommand{\Dn}{{\cal D}}
\newcommand{\Dndagger}{{\cal D}^{\dagger}}
\newcommand{\Dnote}{{\cal D}_0}
\newcommand{\Dnotedagger}{{\cal D}_0^{\dagger}}

\section{Symmetry for Jost functions}
So far we have defined  regular solutions  \(\varphi^n_\pm\) and
 Jost solutions \(f^n_\pm\)  and expressed the spectral density
by  Jost functions.
In this section, we investigate about the one-to-one correspondence
for these solutions and show the
contribution to \(d\ln Z/ d\alpha\) from the nonzero states vanish.

We first prove  a set of relations which hold for the solutions;
\bea \psi^{n+1}_-( k,r)
& = & {1\over k} \left( -{d\over dr} + { n+ 1/2 \over r}
- {eA_\theta\over r } \right) \psi^n_+ ( k,r), \label{eq:susy3}\\
\psi^n_+ (k, r)  & = & {1\over k} \left( {d\over dr} + { n+ 1/2 \over r}
- {eA_\theta\over r } \right) \psi^{n+1}_- (k, r), \label{eq:susy4}\eea
where \(\psi^n_\pm\) denotes both \(f^n_\pm\) and \(\varphi^n_\pm\)
collectively.
These relations are the partial wave versions of
the relations (\ref{eq:susy1})--(\ref{eq:susy2}); note
the differential operators \(D\) and \(D^\dagger\),
(\ref{eq:4.5})--(\ref{eq:4.6}), connect only the partial waves whose angular
momentum differs by one.
Also, these relations holds for any complex values of \(k\)
at which the solutions are finite and well-defined.
In this sense, they are extension of  (\ref{eq:susy1})--(\ref{eq:susy2})
to complex values of energy.

We specifically prove (\ref{eq:susy3}) for \(\Regp kr\) and \(\Regm kr\);
the others can be shown in quite similar way.
Let us use notations
\bea \Dndagger  =  -{d\over dr} + { n+ 1/2 \over r}
- {eA_\theta\over r } =  \Dnotedagger - {eA_\theta\over r } \\
\Dn  = {d\over dr} + { n+ 1/2 \over r} - {eA_\theta\over r } =
\Dnote - {eA_\theta\over r }. \eea
Note a set of formula,
\bea {1\over k} \Dnotedagger \freeu nkr  &=&  \freeu {n+1}kr, \qquad
{1\over k}\Dnote \freeu {n+1} kr = \freeu nkr, \nonumber\\
{1\over k} \Dnotedagger \freew nkr  &=&  \freew {n+1}kr, \qquad
{1\over k}\Dnote \freew {n+1} kr = \freew nkr,\label{eq:u2}\eea
which apply to the free solutions defined by
(\ref{eq:u})--(\ref{eq:w}).
The equations (\ref{eq:reg1}) for \(\Regp kr\) and \(\Regm kr\)
in abbreviated but manifest notations are
\bea
\varphi_+ & = & u^n + {1\over k^2 - \Dnote\Dnotedagger}V_+\varphi_+,
\label{eq:ab1}\\
\varphi_- & = & u^{n+1} + {1\over k^2 - \Dnotedagger\Dnote} V_- \varphi_-.
\label{eq:ab2}
\eea
The potentials \(V_+\) with the angular momentum \(n\)
and \(V_-\) with \(n+1\) are also
written as
\bea V_+ & = & - \Dnote {\bar A} - {\bar A} \Dnotedagger + {\bar A}^2 \\
 V_- & = & - \Dnotedagger {\bar A} - {\bar A} \Dnote + {\bar A}^2,
\eea where \( {\bar A} \equiv e A_\theta/r \).
Using explicit integral expression for \( 1/(k^2 - \Dnote\Dnotedagger)\)
in terms of \( u^n \) and \(w^n\) and applying the formulas
(\ref{eq:u2}),
we verify
\be \Dnotedagger {1\over k^2 - \Dnote\Dnotedagger} =
{1\over k^2 - \Dnotedagger\Dnote}\Dnotedagger \label{eq:G1}\ee
as long as it operates on a function, say \(h(r)\),
that is regular enough at \(r = 0\) so that
\be \lim_{r\rightarrow 0} \freew {n+1} kr h(r) = 0. \label{eq:r0}\ee
This condition is satisfied for the potentials derived by (\ref{eq:potential})
with \(A_\theta\) that satisfies the regularity condition at \(r=0\)
which we have mentioned in section 4.
Similarly, we get
\be \Dnote {1\over k^2 - \Dnotedagger\Dnote} =
{1\over k^2 - \Dnote\Dnotedagger}\Dnote. \label{eq:G2}\ee

Now applying \(\Dnotedagger\) to (\ref{eq:ab1}) and using (\ref{eq:G1}),
we get
\bea  \Dnotedagger \varphi_+ &=& \Dnotedagger u^n +
{1\over k^2 - \Dnotedagger\Dnote}\left[ V_- \Dnotedagger
+ \left( k^2 - \Dnotedagger \Dnote \right) {\bar A} \right.\nonumber\\
&& - \left. {\bar A}\left( k^2 - \Dnote \Dnotedagger \right)
 -  {\bar A}^2 \Dnotedagger
+\Dnotedagger {\bar A}^2 \right] \varphi_+, \eea
where we have used
\be \Dnotedagger V_+ =  V_- \Dnotedagger - \Dnotedagger\Dnote {\bar A}
+ {\bar A} \Dnote\Dnotedagger - {\bar A}^2 \Dnotedagger
+ \Dnotedagger {\bar A}^2. \ee
Thus \(\Dndagger \varphi_+ \) obeys
\bea  \Dndagger \varphi_+ & = & k u^{n+1} +
{1\over k^2 - \Dnotedagger\Dnote}\left[ V_-\Dnotedagger
- {\bar A}\left( k^2 - \Dnote \Dnotedagger \right) \right] \varphi_+
\nonumber\\
&&- {1\over k^2 - \Dnotedagger\Dnote}\left[ {\bar A}^2 \Dnotedagger
- \Dnotedagger {\bar A}^2 \right] \varphi_+. \eea
Iterate \(\varphi_+\) in the second term by (\ref{eq:ab1}) and get
\bea
\Dndagger \varphi_+ & = & k u^{n+1} + {1\over k^2 - \Dnotedagger\Dnote}  V_-
k u^{n+1} \nonumber\\
&&+ {1\over k^2 - \Dnotedagger\Dnote} V_- {1\over k^2 - \Dnotedagger\Dnote}
\left[  V_-\Dnotedagger
- {\bar A}\left( k^2 - \Dnote \Dnotedagger \right) \right] \varphi_+
\nonumber\\
&& - {1\over k^2 - \Dnotedagger\Dnote}  V_-
{1\over k^2 - \Dnotedagger\Dnote}
\left[ {\bar A}^2 \Dnotedagger
- \Dnotedagger {\bar A}^2 \right] \varphi_+.
\eea
Continue the similar iterations and get
\be \Dndagger \varphi_+ = k u^{n+1} +
\sum_{m=1}^{\infty}
\left( {1\over k^2 - \Dnotedagger\Dnote}  V_- \right)^m k u^{n+1}. \ee
This series coincides what is generated by (\ref{eq:ab2}) up to a factor \(k\);
thus we get (\ref{eq:susy3}).

The proof for \(\Jostp kr \) and \( \Jostm kr \) is quite similar.
The equations (\ref{eq:G1}) and (\ref{eq:G2}) holds as long as they operate
on a function \(h(r)\) that obeys
\be \lim_{r\rightarrow \infty } \freeu {n+1} k{r} h(r) = 0, \ee
which is also satisfied in the problem we are considering.

The  relation (\ref{eq:susy3})--(\ref{eq:susy4}) lead very
critical result.
The contribution of the nonzero energy states
to \(d\ln Z/ d\alpha \) is given by
the difference between the spectral densities of \(\Hp\) and \(\Hm\).
According to (\ref{eq:SDinJ}), it reads
\be \rho_+ (E) - \rho_-(E) = {i \over 2\pi} \sum^\infty_{-\infty}
{1\over 2k}{d\over dk} \left[ \ln{\Jostfp k \over\Jostfm k }
- \ln{\Jostfp { - k }\over\Jostfm { -k} }  \right]
\ee
with \( k = \sqrt E\).
On the other hand, because of (\ref{eq:susy3})--(\ref{eq:susy4}), we see
\bea &&  \WronskianOf{ \Jostm kr}{\Regm kr} \nonumber\\
& = & {1\over k^2}\left[{\cal D}^\dagger \Jostp kr {\cal D}{\cal D}^\dagger
\Regp kr - {\cal D}{\cal D}^\dagger \Jostp kr {\cal D}^\dagger
\Regp kr\right] \nonumber\\
& = & \WronskianOf{\Jostp kr}{\Regp kr}, \eea
i.e.
\be \Jostfm k = \Jostfp k. \ee
Thus the contribution of nonzero energy states to \(d\ln Z/ d\alpha \) vanish.

\section{Contribution around \(E = 0\)}
We have seen that \(d\ln Z/d\alpha\) gets contribution only from
the zero energy states, i.e., the integral over C\('\),
(see Eq.~(\ref{eq:ad})).
To know how the resolvents behaves at \(E \sim 0\) is essential to evaluate
the integral.
Due to the \(1/r^2\)-tail in the potential, the analysis is quite difficult.
Instead, we attempt to estimate \(d\ln Z/d\alpha\) by an expansion
in \( 1/|m|^2 \): the \(E\)-integral is now carried out over finite region;
thus we expect the
expansion of \(|m|^2 /(E + |m|^2) \) in the integrand in \(1/|m|^2\) results
in a decent expansion, which is at least asymptotic to \(d\ln Z/d \alpha\) as
\(|m|\) goes to infinity.
Let us define
\be
S_k \equiv \int d^2\!x \lim_{  y \rightarrow  x}
{1\over 2\pi i} \oint_{C'} dE (-E)^k
\left[ \Rp Exy - \Rm Exy \right]
\label{eq:Sk}\ee
and assume they are finite.
The expansion is then given by
\be
{ d\ln Z \over d\alpha } = i \left( S_0 + {1\over |m|^2} S_1 +
{1\over |m|^4} S_2 + ...\right).
\ee

In order to get the coefficients \( S_k \), we look into
\be
\Delta (\tau) \equiv \int d^2\!x \Ano(\tau, x).
\ee
Recalling Eq.~(\ref{eq:AinR}), we notice that \(\Delta(\tau) \) also get the
contribution only from integral over C\('\)
and thus has the expansion in \(\tau\) with the coefficients \(S_k\).
The advantage of considering \(\Delta (\tau) \) is that
the asymptotic expansion
of \(\Ano( \tau, x)\),
\be \Ano(\tau, x) \sim \Ano_0(x) + \tau \Ano_1(x) + \tau^2 \Ano_2(x) + ... ,
\label{eq:EofA}\ee
can be evaluated by a straight forward way\footnote{%
The interplay between the index of Fredholm operator and the asymptotic
expansion of its heat kernel is described in \cite{Gilkey:1984}};
once we get \(\Ano_k (x) \), \(S_k\) are given by integrating them,
\be S_k =  k! \int d^2\!x \Ano_k(x). \label{eq:SbyA}\ee

We evaluate the coefficients \(\Ano_k(x)\) using the explicit expression
for \(\Ano(\tau, x)\) \cite{Fujikawa:1979,Fujikawa:1984},
\be
\Ano (\tau,x) =  \lim_{y\rightarrow x}
\int {d^2p \over (2\pi)^2} \left( e^{ - \tau \Hp } -
e^{ - \tau \Hm } \right) e^{i p \cdot (x-y)}.
\ee
We regard the operation of \(e^{ -\tau \Hpm }\) is calculated  by the
corresponding power series.
To take the limit \(y\rightarrow x\), we move the factor \(e^{ ip\cdot x }\)
on the right side of the operators to left.
Every time one commutes the partial derivative \(\partial_\mu\)
with \(e^{ ip\cdot x }\),
we changes it with \(\partial_\mu + i p _\mu \).
Scaling  the momentum by \(\sqrt\tau\), we get
\bea
\lefteqn{  \Ano(\tau, x)  = }\nonumber\\
& & {1\over\tau} \int { d^2p\over (2\pi)^2}
e^{- p^2} \sum_{n=0} {1\over n!} \left\{ \tau \left[
 { 2 i p\cdot( \partial -ieA )\over \sqrt\tau}
 + ( \partial -ieA )^2 + e F_{12}
\right]  \right\}^n  \nonumber\\
&& - ( F_{12} \rightarrow -F_{12} ).\label{eq:Ano} \eea
The expansion (\ref{eq:EofA}) is given by picking  up terms
with the same powers in \(\tau\).

We have evaluated the series explicitly up to \(\Ano_2\)
and got
\bea \Ano_0( x ) & = & {1\over 2\pi}  e F_{12},\nonumber\\
\Ano_1(x)  & = & {1\over 12\pi} \partial_\mu^2 e F_{12},  \nonumber\\
\Ano_2(x)  & = & {1\over 120\pi} \partial_\mu^2 \partial_\nu^2 e F_{12}.
\eea
Note all these terms must  be  gauge invariant as well as having parity odd.
It is further interesting to note that the results are total divergence:
even under the requirement of the gauge invariance and the parity,
the term proportional to \((F_{12})^3\)  was possible for \(\Ano_3\);
these terms cancel out in the explicit calculation.

The coefficients \(S_k\) are now calculated by (\ref{eq:SbyA}).
The leading term is
\be S_0 = \nu, \ee
i.e., the winding number of background field.
This is  exactly the same way as in four dimensional QCD.
Since \(F_{12}\) approaches zero faster than \(1/r^3\), the
second and third correction terms are zero.

\section{Summary and Discussion}
In summarizing, we have evaluated the chiral phase dependence of free
energy \(\ln Z\) in two dimensional spherically symmetric background.
We divided up the contribution into two; one from the nonzero energy states
and one from  zero energy states regularizing the later by
the contour integral over C\('\).
The contribution from the nonzero energy states is shown to be zero.
In other words, the one-to-one correspondence is exact.
As for the zero energy states contribution, we proposed the expansion
in \(1/|m|^2\).
The coefficients are given explicitly by the behavior of the resolvents at
\(E\sim 0\), see Eq.~(\ref{eq:Sk}).
We have evaluated the coefficients up to the second power of  \(1/|m|^2\).
We found  the \(\alpha\)-dependence of the free energy  is given
by the winding number of the background U(1) gauge field and
corrections to this relation is zero up to \(1/|m|^4\).

The expansion for \(\Ano(\tau, x)\) also implies that the expectation value of
the axial U(1) current can be given by the expansion of \(1/|m|^2\).
Comparing  our result with the anomaly equation for
the axial U(1) current
\be
m \bar\psi(1+\sigma_c )\psi - m^* \bar\psi(1-\sigma_c )\psi =
{ie\over\pi}F_{12} - \partial_\mu \left( \bar\psi\sigma_\mu\sigma_c\psi\right)
, \ee
we get
\be \partial_\mu \left\langle \bar\psi\sigma_\mu\sigma_c\psi\right\rangle
\sim (-i) \left( { \partial_\mu^2 e F_{12} \over 6\pi |m|^2 } +
{ \partial_\mu^2 \partial_\nu^2 e F_{12}\over 30\pi |m|^4 } + ...\right).
\ee

Finally, it is worth mentioning the interplay between nonzero winding
number and \(1/r^2\)-tail in the potential.
The existence of the \(1/r^2\)-tail means the violation of the Levinson theorem
in the scattering problem \cite{Chadan.et.al:1989}.
The violation indicates that the total number of states
in the spectra of \(\Hpm\) are not the same.
This difference causes the nontrivial
\(\alpha\)-dependence of the Jacobian in the path integral measure
when we rotate \(\alpha\) away from the mass term in the lagrangian.

\medskip
\begin{center}
\bf Acknowledgement
\end{center}
 This work is supported in part by the U.S. Department of Energy under contract
No.~DE-AT03-87ER40327.

\appendix
\section{Appendix} \label{app:consistency}
We verify in this appendix that the definition (\ref{eq:dens}) for
the state density \( \SD Ex \) gives correct  results  for  hamiltonians
which describe a particle moving freely in \( d\)-dimensional space.
The hamiltonian is
\be H = - \left( \partial \over \partial x_\mu \right) ^ 2 \ee
with an unit \( 2m = 1 \) for the particle mass \( m\); \(\mu\) runs
from 1 to \(d\).
According to the standard box quantization, the number of quantum states
per unit momenta and coordinates, is
\be { dN \over \prod_\mu dp_\mu dx_\mu } = {1\over (2\pi)^d } \ee
in an unit \( \hbar = 1 \).
The state density \(\SD Ex\), the number of states per unit energy
and unit coordinates is thus
\be \SD E x = {1\over (2\pi)^d} \int d^d p\, \del E  {p^2}
= { s(d) \over 2 (2\pi)^d } \left( \sqrt E \right)^{d-2}, \label{eq:a1}\ee
where \(s(d)\) is volume of unit sphere S\(^{d-1}\),
\(s(1) = 2, s(2) = 2\pi, s(3) = 4\pi, ... \)
(\(s(1) \) is  the number of points that satisfy \(p^2 = E\).)

For \( d= 1\), the resolvent \(\R Exy \) is
\be \R Exy = {1\over 2\sqrt {-E}} e^{-\sqrt{-E}\,|x-y|}, \ee
where we  choose a cut of \(\sqrt{-E}\) along positive real \(E\)-axis.
Then,
\be \SD Ex = \lim_{\epsilon \rightarrow +0}
{1\over 2\pi i} \left( {1\over 2\sqrt{-E -i\epsilon}} -
{1\over 2\sqrt{-E +i\epsilon}} \right) = {1\over 2\pi \sqrt E}. \ee
Similarly, for \( d=2\),
\be \R Exy \sim - {1\over 2\pi} \ln \left( \sqrt{-E} \, | x -  y |\right)
\quad\mbox{for }  |x - y| \ll  1/\sqrt {|E|}\ee
and
\bea \SD Ex & = & \lim_{\epsilon \rightarrow +0}
{1\over 2\pi i} \left[ - {1\over 2\pi} \ln \left( \sqrt{-E -i\epsilon }
\right) + {1\over 2\pi} \ln \left(
\sqrt{-E + i\epsilon } \right) \right] \nonumber\\
& = & { 1\over 4\pi}. \eea
For \( d = 3\),
 \be \R Exy = {1\over 4\pi | x - y| } e^{ - \sqrt{-E }
\, | x -  y |}, \ee
 and
\bea \SD Ex & = & \lim_{y\rightarrow  x} \lim_{\epsilon \rightarrow +0}
{1\over 2\pi i} {1\over 4\pi |x -  y| }
\left[ e^{ - \sqrt{- E -i\epsilon}\,| x -  y| } -
e^{ - \sqrt{-E + i\epsilon}\, | x -  y| } \right] \nonumber\\
& = & {\sqrt E \over 4\pi^2 }. \eea
All these results are consistent with (\ref{eq:a1}).
Note that for \( d \ge 2 \), \(\R E xy \) diverges as \( y \rightarrow x\).
Thus in order to avoid the ill-definedness,
we have to calculate discontinuities
at the cut first before taking the limit \(y\rightarrow x\).

\section{Appendix} \label{app:density}
In this appendix we show (\ref{eq:SDinJ}). To see this, we calculate
\be I(k) \equiv  \left.{d \ln F^n \over dk}\right|_k
  + \left.{d \ln F^n \over dk}\right|_{-k^*}
= {\dot \Jostf nk \over \Jostf nk}
+ {\dot \Jostf n{-k^*} \over \Jostf n{-k^*}}
\ee
and take  a limit that \(k\) approaches to real positive
values, \( k \sim |k| + i\epsilon \),
where `dot' denotes derivative with respect to \(k\).
Using the definition of the Jost function (\ref{eq:jostf}),
we obtain
\be I(k) = I_1(k,r) + I_2 (k,r), \ee
where
\bea
I_1(k,r) & \equiv &  {1\over \Jostf nk} \WronskianOf { \Jost nkr}{\dot \Reg
nkr}
\nonumber\\
&&+ {1\over \Jostf n{-k^*} } \WronskianOf { \Jost n{-k^*} r}{\dot \Reg
n{-k^*}r}
\eea
and
\bea
I_2(k,r) & \equiv &  {1\over \Jostf nk} \WronskianOf {\dot \Jost nkr}{\Reg nkr}
\nonumber\\
&&+ {1\over \Jostf n{-k^*} } \WronskianOf {\dot \Jost n{-k^*} r}{\Reg
n{-k^*}r}.
\label{eq:I2}\eea
Since
\be \dot\Reg n {-k^*}r = (-)^{\absn+ 1} \left[ \dot \Reg nkr \right]^* \ee
and \(\dot \Reg nkr\) becomes  real as \(k\) approaches to the real value,
\be I_1(k, r) \rightarrow
{2i \over \Jostf nk \Jostf n {-k} } \WronskianOf {\Reg nkr}{\dot \Reg nkr }
\quad \mbox{as } \epsilon \rightarrow 0,\ee
where we have used (\ref{eq:reg3}) and (\ref{eq:RbyJ}).
Noting that
\( \dot \Reg nkr \) obeys an equation
\be \left[ - { d^2 \over dr^2 } + {1\over r^2}\left( n^2 - {1\over 4} \right) +
V(r) \right] \dot \Reg  nkr = k^2 \dot \Reg nkr + 2 k
\Reg nkr, \ee
which is given by differentiating (\ref{eq:scte}),
we obtain
\be {d\over dr} \WronskianOf{ \Reg nkr}{ \dot \Reg nkr} = -2k \left[ \Reg nkr
\right]^2. \ee
Integrating this equation with the boundary condition
\be \lim_{r\rightarrow 0} \WronskianOf{\Reg nkr}{\dot \Reg nkr } = 0 \ee
(see equation (\ref{eq:Reg23})),  we get
\be
I_1(k,r) \rightarrow
{- 4ik \over \Jostf nk \Jostf n {-k} } \int_0^r dr' \left[
\Reg nk{r'} \right]^2 \quad \mbox{as } \epsilon \rightarrow 0. \ee

On the other hand, \(I_2(k,r)\) becomes independent from  specific potential
of the scattering problem  in the limit of \(r\rightarrow \infty\),
which can be seen in the following way.
\(\Jost nkr \) approaches to the free solution and thus
\be \dot\Jost nkr \rightarrow
\dot \freew nkr \quad\mbox{as } r\rightarrow \infty .\label{eq:f1}\ee
\(\Reg nkr \)  consists of exponentially decreasing component
\( e^{-\epsilon r}\)  and increasing component \(e^{\epsilon r}\) at the
complex value of \(k\).
Since \(\dot \Jost nkr\)  decreases exponentially, only the increasing
component can remain in (\ref{eq:I2}) as \( r\rightarrow \infty \).
The  Jost function is the coefficient of the increasing component
in the expansion of \(\Reg nkr \).
Because of (\ref{eq:Wron}), we can specifically write it as
\be \Reg nkr \rightarrow  \Jostf nk \freeu nkr
\quad\mbox{as } r\rightarrow \infty ,\label{eq:f2}\ee
where we are only interested in the increasing component in \(\freeu nkr\).
Plugging (\ref{eq:f1}) and (\ref{eq:f2}) into (\ref{eq:I2}), we see
that the potential dependence, the Jost function \(\Jostf nk\),
cancels in numerator and denominator.

\(I(k)\) for real \(k\) is now written by taking \(r\rightarrow \infty\)
in  \(I_1(k,r)\) and \(I_2(k,r)\) as
\be I(k) = \lim_{r\rightarrow \infty} \left(
{- 4ik \over \Jostf nk \Jostf n {-k} } \int_0^r dr' \left[
\Reg nk{r'} \right]^2 + I_2(k,r) \right), \ee
and the contribution of the second term is potential-independent.
Note the first term is related to the integral of \(\SD Ex \) over \(x\).
Further note \(\Jostf nk = 0\) and \(I(k) = 0\) for the free potential;
thus \(I_2(k,r)\) should satisfy
\be I_2(k,r) \rightarrow  -4ik \int_0^r dr' \left[\freeu nk{r'} \right]^2
\quad \mbox{as } r\rightarrow \infty, \ee
which is related to spectral density for free potential.
Thus \(\delta\rho(E)\), the deviation of spectral density from one for free
potential, is given by summing \(I(k)\) over all angular momentum \(n\) as
\be \delta\rho(E) = {-1\over 4\pi i k} \sum_{n=-\infty}^{\infty} I(k), \ee
which is (\ref{eq:SDinJ}).

\newpage

\begin{center}
\bf Figure captions
\end{center}
\begin{description}
\item[Fig.~1.]  The integral contour C\('\) on the complex \(E\)-plane
with a cut along the real positive axis.
\end{description}

\end{document}